\documentclass[aps,prd]{revtex4}
\usepackage{graphicx}
\begin{document}

\title{Understanding $K/\pi$ ratio distribution in the mixed events}
\author{C.B. Yang$^{1,2}$, X. Cai$^1$ and Z.M. Zhou$^3$}
\affiliation{$^1$ Institute of Particle Physics, Hua-Zhong Normal University,
Wuhan 430079, P.R. China\\
$^2$ Max-Planck-Institut f\"ur Physik, F\"ohringer Ring 6, D-80805 M\"unchen, Germany\\
$^3$ Physics Department, Hua-Zhong University of Science and Technology, Wuhan 430074,
P.R. China}
\date{\today}

\begin{abstract}
The event mixing method is analyzed for the study of the event-by-event $K/\pi$ ratio
distribution. It is shown that there exists some correlation between the kaon
and pion multiplicities in the mixed events. The $K/\pi$ ratio distributions in the
mixed events for different sets of real events are shown. The dependence of the
distributions on the mean $K/\pi$ ratio, mean and variance of multiplicity distribution
in the real events is investigated systematically. The effect of imperfect particle
identification on the $K/\pi$ ratio distribution in the mixed event is also considered.

{{\bf PACS number(s)}: 12.40.Ee, 25.75.-q, 13.60.Le}
\end{abstract}

\maketitle

\section{Introduction}

The strangeness production in high energy heavy ion collisions is an important
topic. Because there is no strangeness content in the initial states of the collisions,
the strangeness in the final states is a result of the complex interactions in the
collisions. It has been proved that the strangeness in the final states is sensitive
to the dynamics of the process. The enhanced strangeness production in
ultra-relativistic heavy ion collisions may be caused by:
(i) equilibrated gluon rich plasma phase \cite{shuryak}; (ii) baryon-junctions \cite{vance};
(iii) diquark breaking and sea-diquark \cite{capella}; (iv) strong color fields \cite{bleicher}
etc. In particular, the enhanced production may be related
to the formation of the long searching deconfined quark-gluon plasma (QGP) \cite{pr142} and
can shed light on the time scale of chemical freeze-out and therefore carry information
about the early stage of a heavy ion collision \cite{lpc94}.

In recent years, with the advent of large acceptance detectors, event-by-event (EBE) analysis
\cite{zpc54,epjc6,appb29,plb435,plb439,plb459,epjc8} has been performed in many investigations.
The philosophy of EBE physics is based on following speculation: Although
the conditions for QGP production may be reached in every event, the fact that a phase
transition from QGP state to hadronic final state is a critical phenomenon implies that
deconfined QGP state may occur only in a very small sub-sample of events. So the fluctuations
accompanying the phase transition will, in effect, be averaged out in the conventional ensemble
analyses. The EBE analysis searches for fluctuations of observables at the
event level, so it can be used to select interesting or anomalous event candidates
\cite{nato94} with specific dynamical properties. So the EBE analysis can provide dynamical
information which cannot be obtained from the traditional inclusive spectra.

To estimate the amount of dynamical contributions to the distribution of EBE observables, the
method of event mixing has been used to create reference events which may contain only
statistical fluctuations and have the same inclusive (momentum and etc.) spectra as the original
ones. To create mixed events one first sets up a track pool (``event-mixer'') and inserts many
(a few hundred or more) real events into the mixer. Then, by selecting particles from different
original events in the pool one can get a non-real (mixed) event. In the process of constructing
a mixed event only one track in any original event is allowed to be used.
The number of particles in a mixed event
is a random integer satisfying the multiplicity distribution determined from the original
real events in the filled pool. Thus, the mixed events are random samples from the same
inclusive track population as the real events, but there is no internal momentum correlations
in them. So, the mixed events may be useful in the study of correlations due to some
conservation laws, such as energy-momentum conservation law, since the
constraint of such conservation laws plays no role in the mixed events.

Recently the EBE analysis is used in experiments to analyze flavor (strangeness)
fluctuations in high energy heavy ion collisions \cite{gr97,fs99,croland99}.
In one of our earlier papers \cite{yang2001} we tried to calculate the trivial $K/\pi$ ratio
distribution for events having statistically independent multiplicity distributions for
kaons and pions. For that pure statistical case, the $K/\pi$ ratio distribution was shown to be
approximately Gaussian, and its width depends on the shape of the distributions of kaon and pion
multiplicities. That result is the starting point for investigating any dynamical effect in the
production of strange kaons, though it is not for the real situation in physical problems.
In real experimental case, the multiplicity distributions for kaons and pions may be not
independent, and the mixed events were used in \cite{gr97,fs99,croland99} to estimate the
statistical contributions to the ratio distribution. In calculating
the EBE $K/\pi$ ratio, the only things relevant are
the numbers of kaons and pions in a real and/or mixed events. Unlike in the study of mean
transverse momentum fluctuations, where the inclusive transverse momentum distributions
for real and mixed events are the same, the inclusive distributions of kaon and pion
multiplicities are not necessarily the same for real and mixed events. One can see this point
clearly in Sect. \ref{Sec:ff}. To the best of our knowledge, it is not known yet what will be
the $K/\pi$ ratio distribution in the mixed events before event mixing is actually
performed when the ratio distribution in real events is given. Such knowledge
is important for us to understand the true effect of event mixing.
It will also speed up the real data analysis and save a lot of computing resources.
We should investigate theoretically on which quantities in the real events
the $K/\pi$ ratio distribution in the mixed events depends.
The first question one should ask is: What is the dependence of the
ratio distribution in the mixed events on that in the original events? If such dependence
is strong, the mixed events can hardly be used to estimate the dynamical contribution
in the real data.

In this paper, we try to study systematically the effect of event mixing on the EBE $K/\pi$
ratio distribution. By $K/\pi$ we denote the ratio of positively charged kaons to pions in the
events. For the sake of simplicity, we assume that there are only pions and kaons in the
phase space under study. This paper is organized as following. After some general
discussion in Sect. \ref{Sec:gen}, we will start in Sect. \ref{Sec:ff} from the simplest
case by assuming that the multiplicity and $K/\pi$ ratio are both fixed in the
original events. For this case, the effect of event mixing on the $K/\pi$ ratio distribution
can be treated with rigorously and analytically. Then we will study the influence of
$K/\pi$ ratio distribution in real events on the event mixing in Sect. \ref{Sec:fd}.
There it is assumed that all the real events have common fixed multiplicity but different
EBE $K/\pi$ ratios. Sect. \ref{Sec:df} will devote to the event mixing for original
events having a non-trivial multiplicity distribution but with a common fixed EBE $K/\pi$
ratio. Sect. \ref{Sec:dd} is for the most general case for mixing of events with
{\em distributed} multiplicity and {\em distributed} $K/\pi$ ratio. In Sect. \ref{Sec:intr}
we will consider the influence of imperfect particle identification on the ratio distributions
in the real and mixed events. Sect. \ref{Sec:con} is for discussion and conclusion.

\section{General discussions}\label{Sec:gen}

We first give some general formula for the calculation of some quantities characterizing
the $K/\pi$ ratio distribution.

Suppose the probability for an event (real or mixed) to have $N-k$ pions and $k$ kaons
is $P(N,k)$. Define a generating function
\begin{equation}
G(x,y)=\sum_k\sum_\pi x^ky^\pi P(\pi+k,k)\quad,
\label{Eq:gen}
\end{equation}

\noindent then we have a global $K/\pi$ ratio $R_g$ as
\begin{equation}
R_g\equiv {\langle k\rangle\over \langle\pi\rangle}=G_x^\prime(1,1)/G_y^\prime(1,1)\quad,
\label{Eq:global}
\end{equation}

\noindent where the subscript together with a prime represents derivative of the generating
function with respect to the variables. In the expression,
$\langle\cdots\rangle$ represents average over all possible configuration.

In an EBE analysis, one measures EBE ratio $r_e\equiv k/\pi$ for every event and
observes the distribution of the EBE ratio. For such an analysis,
one can get
\begin{equation}
R_e\equiv \langle k/\pi\rangle=\left.{d\over dx}\right|_{x=1}\int_0^1 dy{G(x,y)-G(x,0)
\over y} \label{Eq:gl1}
\end{equation}
\begin{equation}
\langle k^2/\pi^2\rangle=\left.\left({d^2\over dx^2}+{d\over dx}\right)\right|_{x=1}
\int_0^1 dy {G(x,y)-G(x,0)\over y}\ln{1\over y}\quad,\label{Eq:gl2}
\end{equation}

\noindent
then one can get the width of the EBE $K/\pi$ ratio distribution
as $\sigma=\sqrt{\langle k^2/\pi^2\rangle-\langle k/\pi\rangle^2}$.  In deriving last
equation condition $1-P(N,N)\simeq 1$ is used, which is quite a good approximation for
current experimental analysis.

\section{Mixing of events with {\em fixed} multiplicity and {\em fixed} $K/\pi$ ratio}
\label{Sec:ff}

We begin to study the effect of event mixing from the simplest case. We assume that every
real original event has the same fixed numbers of kaons and pions. Then the $K/\pi$ ratio
in the original events is fixed, so does the total multiplicity $N$. We denote in this paper
such a case by FF ({\it fixed} multiplicity and {\it fixed} $K/\pi$ ratio). No chemical
fluctuations in the real events are assumed.

Then, we consider the mixing of the original events. By construction, the mixed events
have the same total multiplicity $N$ as the original ones. When a mixed event is constructed,
$N$ particles are selected from the mixer, and one picks up one and only one particle
randomly from tracks of a randomly chosen original event. Since the numbers of kaons and pions
are assumed fixed in the original events, the selected particle from an original event can
be a pion or a kaon with definite probability. Assume the multiplicity ratio of kaons to pions
is $R$, the probability for the chosen particle to be a kaon is $R/(1+R)$, and that
for a chosen particle to be a pion is $1/(1+R)$. Then, the probability for a mixed event
to have $k$ kaons and $\pi=N-k$ pions is prescribed by binomial distribution
\begin{equation}
P_{\rm FF}(N,k)=\left(\begin{array}{c} N\\ k\end{array}\right)
\left({R\over 1+R}\right)^k\left({1\over 1+R}\right)^{N-k}\quad.
\label{Eq:ffp}
\end{equation}

\noindent It is obvious that
the multiplicity distributions for kaons and pions in the mixed events
are determined by the total (kaons plus pions) multiplicity $N$ together and the $K/\pi$
ratio $R$ in real events. One sees that the inclusive kaon (or pion) multiplicity distribution
in the mixed events is very different from that in original events.

\begin{figure}\centering
\includegraphics[width=0.5\textwidth]{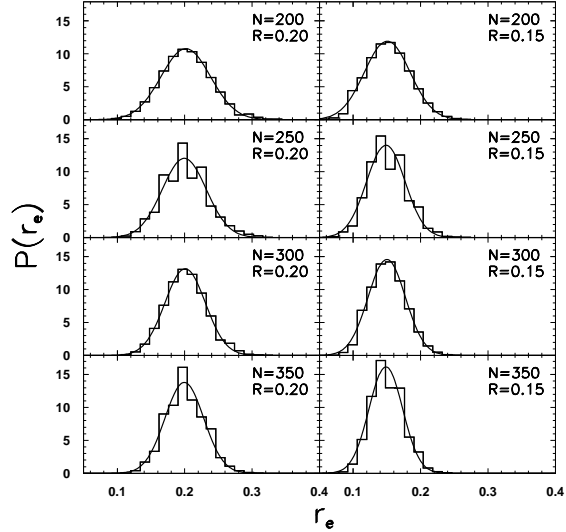}
\caption{Distribution of EBE $K/\pi$ ratios in the mixed events from a set of
original events with {\it fixed} multiplicity and {\it fixed} kaon contents.}
\label{Fig:1}
\end{figure}

The EBE $K/\pi$ ratio distribution in the mixed events together with its dependence on
the multiplicity $N$ and the original ratio $R$ can be obtained from Monte Carlo
simulation. To get a mixed event one can generate $N$ random numbers. If a random number
is less than $R/(1+R)$, we say that a kaon is produced. Otherwise, a pion is thought
to be produced. If the total kaon number in a mixed event is $k$, one can get a $K/\pi$ ratio
for the mixed event $r_e=k/(N-k)$. In Fig. \ref{Fig:1} the normalized $r_e$ distributions
$P(r_e)$ are shown for $R$=0.15 and 0.2 as examples. For each case 5 million
mixed events are generated. For each $R$, four values of $N$, 200,250, 300 and 350,
are taken. Gaussian fits to all the distributions are shown in the same figure.
For fixed $R$, one can see that the width of the distribution decreases with
the increase of $N$. For the same $N$, the width increases with the increase of $R$.

\begin{figure}
\centering
\includegraphics[width=0.5\textwidth]{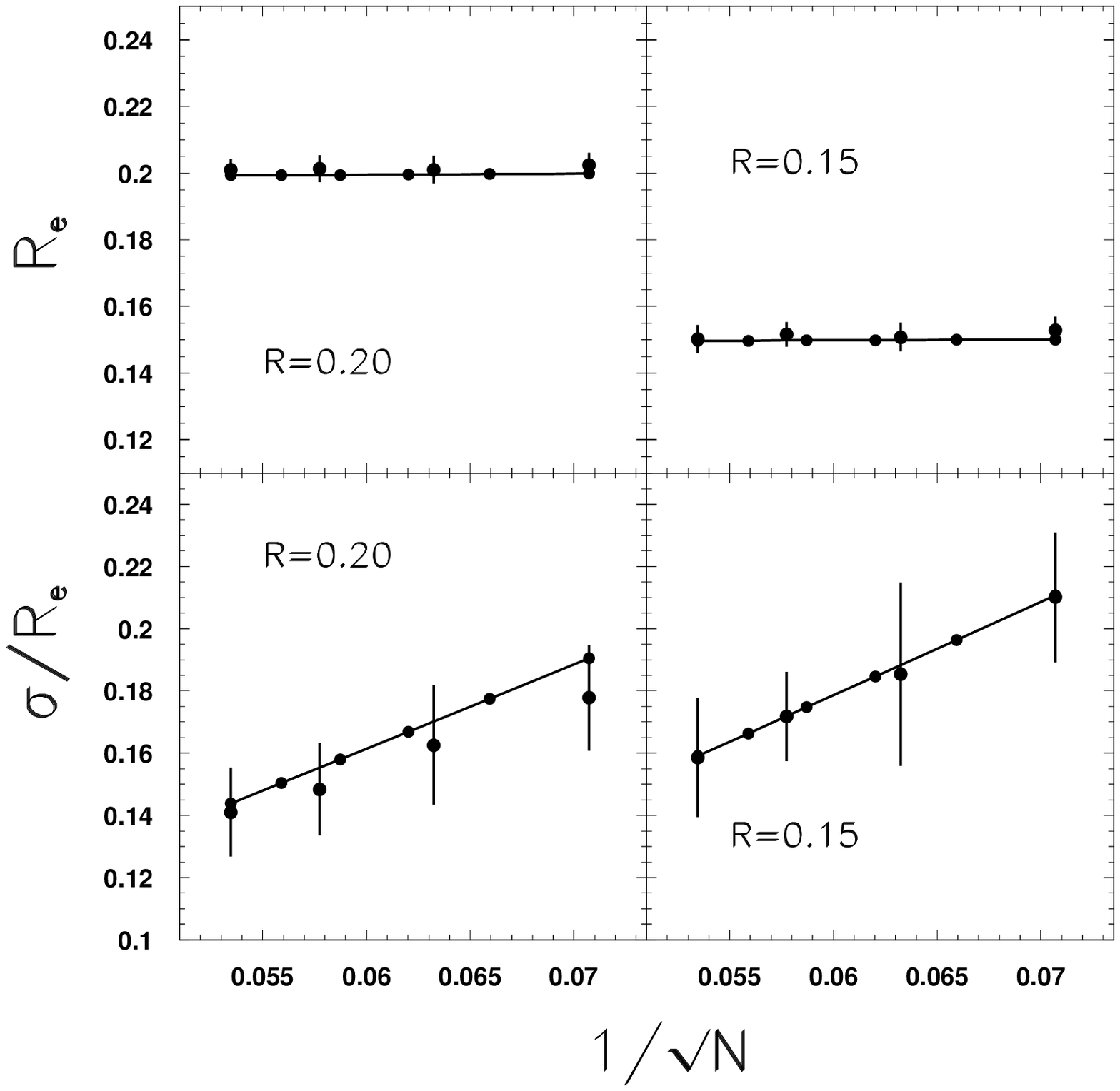}
\caption{Mean values $R_e$ and relative widths $\sigma/R_e$ of the EBE $K/\pi$ ratio
distributions in the mixed events as functions of the inverse root of the event multiplicity.
The points with error bars are fitted results of Fig. \ref{Fig:1} to Gaussian distributions,
and points linked by curves are from theoretical calculation.}
\label{Fig:2}
\end{figure}

The mean values and widths of the distributions can be obtained from a Gaussian fit to the
histograms in Fig. \ref{Fig:1}, and the fitted results are shown with error bars in
Fig. \ref{Fig:2} as functions of inverse root of multiplicity,
$1/\sqrt{N}$. It seems that finite particle number effect plays the only role
in the mixing process for both $R=0.2$ and 0.15 since $\sigma/R\propto 1/\sqrt{N}$ is quite
a good approximation for the points shown in the figure. The slope for
$R=0.15$ is slightly larger than that for $R=0.20$. The mean and the width of the distributions
can also be calculated from Eq. (\ref{Eq:ffp}). To use our general equations Eqs. (\ref{Eq:gl1})
and (\ref{Eq:gl2}), one needs to have the generating function for distribution
Eq. (\ref{Eq:ffp}).  The generating function for such a distribution is
\begin{equation}
G_{\rm FF}(x,y)=\left({Rx+y\over 1+R}\right)^N
\label{Eq:ffg}
\end{equation}

The calculated results for six multiplicities $N$ are also shown in Fig. \ref{Fig:2} as solid
circles linked by curves. One can find that the mean values from two different ways agree
with each other quite well. But there is slight difference between the widths
from the two ways. The slight difference indicates that the EBE $K/\pi$ ratio distribution
in the mixed events is not a perfect Gaussian. For smaller $R$, the
difference between the results from two ways is smaller.

\section{Mixing of events with {\em fixed} multiplicity and {\em distributed} $K/\pi$ ratios}
\label{Sec:fd}

As the second simple case for investigating effect of event mixing on $K/\pi$ ratio
distribution, we assume that the real events have fixed common multiplicity $N$ but different
strange contents (FD). Thus the real events have a non-trivial $K/\pi$ ratio distribution.

Assume that the $N$ particles in a mixed event are picked up from real events labelled
$i_1, i_2, \cdots,i_N$ and that
the corresponding $K/\pi$ ratio in those events are $r_{i_1}, r_{i_2},\cdots,r_{i_N}$,
respectively. Not losing any generality, we assume that the particles picked from
events $i_1,i_2,\cdots,i_k$ are kaons and others are pions. Since we have no interest
in the order of the original events from which
the kaons are picked up, the probability for such
a configuration is then
\begin{equation}
 \left(\begin{array}{c}N\\ k\end{array}
\right) {r_1\over 1+r_1}{r_2\over 1+r_2}\cdots{r_k\over 1+r_k}{1\over 1+r_{k+1}}\cdots
{1\over 1+r_N}\ .
\end{equation}

\noindent Let the mean $K/\pi$ ratio in the original real events be $R$, and rewrite
$r_i$ as $R+\delta_i$. Since the mixed events are constructed by randomly picking up particles
from randomly chosen original events, the final probability for a mixed event to have $k$ kaons
and $N-k$ pions can be obtained from above expression by averaging it over all
possible configurations. Since $\langle\delta_i^{2m+1}\rangle=0$ for $m$=0, 1, 2, $\cdots$,
one can get
\begin{equation}\label{Eq:fdp}
P_{\rm FD}(N,k)=P_{FF}(N,k)-{\sigma_0^2N\over (1+R)^2}\left[P_{FF}(N-1,k-1)-P_{FF}(N,k)\right]
+O(\sigma_0^4)\quad,
\end{equation}

\noindent where $\sigma_0^2=\langle\delta_i^2\rangle$ is the width squared for the ratio
distribution in the real events. As stated above, the value of $\sigma^2_0$ is determined
by the involved statistical and dynamical contributions in the strangeness production.
 Normally, $\sigma_0^2$ is quite small,
therefore the contribution from higher order terms can be safely neglected. From last equation
one can see that the width of the EBE $K/\pi$ distribution will be slightly
larger than in the case with {\em fixed} multiplicity and {\em fixed}
$K/\pi$ ratio. From this equation the mean and width of $K/\pi$ ratio distribution can
 be calculated. To compare with the calculation, we can also perform a Monte Carlo simulation
for the case. We assume that the distribution of $K/\pi$ ratios in the original events
is a Gaussian with a relative width $\sigma_0/R=0.2$. With the same choices of $R$ and $N$ as
in Fig. \ref{Fig:1}, eight distributions of the EBE $K/\pi$ ratios in 5 million mixed events
are shown in Fig. \ref{Fig:3}. One can compare this figure with Fig. \ref{Fig:1} and find that
the corresponding distributions are almost the same. This is not surprising since
the width $\sigma_0$ used in the simulations shown in Fig. \ref{Fig:3} is only 0.03 for
$R=0.15$ or 0.04 for $R=0.2$. Thus the $\sigma_0^2$ term in Eq. ({\ref{Eq:fdp}) has little
effect on the final results. The mean values and relative widths of the distributions
are also very close to those shown in Fig. \ref{Fig:2} and will not be discussed any more.
The insensitivity of the relative width $\sigma/R_e$ (or the ratio distribution) in the mixed
events to the detail of the ratio distribution of original real events is crucial for detecting
dynamical contributions in experimental data analysis from the event mixing method.
Because the original distribution is a mixture result from dynamical and statistical
contributions and the dynamical correlations in the original events are moved
in the mixed events, the difference between the two distributions from real and mixed events
can be used to quantify the dynamical contribution to the distribution in the real events.
But it should be pointed out that the dynamical contribution can make the width of
the ratio distribution larger or smaller, as have been found in \cite{croland99}.
So, we think it better to use $\sigma^2_{\rm dyn}=
(\sigma_{\rm mixed}-\sigma_0)^2$ to measure the dynamical contribution instead of
$\sigma_{\rm mixed}^2-\sigma_0^2$ in \cite{croland99}.
\begin{figure}\centering
\includegraphics[width=0.5\textwidth]{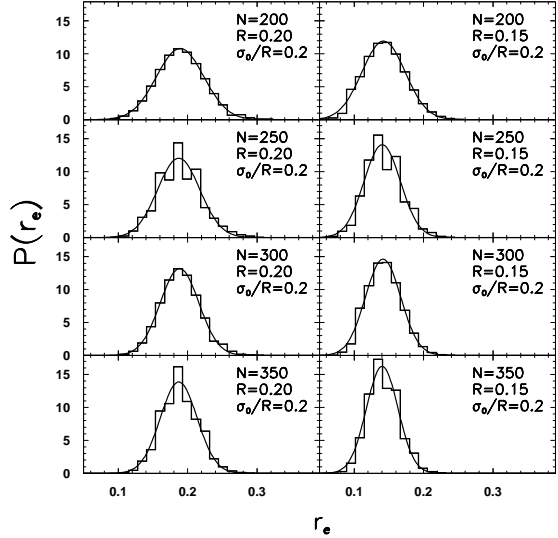}
\caption{Distributions of the EBE $K/\pi$ ratios in the mixed events from original
ones with {\it fixed} multiplicity and {\it distributed} ratios. The $K/\pi$ ratio
distributions in original events are assumed to have a relative width $\sigma_0/R=0.2$
for illustration.}
\label{Fig:3}
\end{figure}

\section{Mixing of events with {\em distributed} multiplicity and {\em fixed} $K/\pi$ ratio}
\label{Sec:df}

In real experiments the event multiplicity is normally not fixed but has a non-trivial
distribution. The multiplicity distribution will certainly affect the $K/\pi$ ratio distribution
in the mixed events, because in Sect. \ref{Sec:ff} we have seen that the ratio distributions
for mixed events with different multiplicities are different. To understand the event
mixing for real experimental events, it is an important step to investigate the mixing of
events with some multiplicity distribution.

Suppose that the multiplicity distribution in the original events is $M(N)$ and that all the
real events have the same fixed $K/\pi$ ratio $R$. We denote this set of original events by
DF. Since the mixed events have the same
total multiplicity distribution as the original ones,
the probability for a mixed event to have $k$ kaon particles and $N-k$ pions is then
\begin{equation}\label{Eq:dfp}
P_{\rm DF}(N,k)=M(N)P_{FF}(N,k)\ .
\end{equation}

From this probability one can get the inclusive kaon (pion) number distribution by summing over
pion (kaon) number. It is clear that the joint probability $P_{DF}(k+\pi,k)$ cannot in general
be equal to the product of the inclusive pion and kaon number distributions. One can see this
point from a simple fact that
\begin{displaymath}
\langle k\pi\rangle=\sum_k\sum_\pi k\pi P_{DF}(k+\pi,k)\neq \sum_k\sum_\pi k
P_{\rm DF}(k+\pi,k)\times \sum_k\sum_\pi \pi P_{DF}(k+\pi,k)=\langle k\rangle\langle\pi\rangle\ .
\end{displaymath}

\noindent
This point can be directly checked even for the case FF with Eq. (\ref{Eq:ffp}). From
that equation one can get
\begin{displaymath}
\langle k\pi\rangle=N\langle k\rangle-\langle k^2\rangle\ ,\quad\quad
\langle k\rangle \langle\pi\rangle=N\langle k\rangle-\langle k\rangle^2\ .
\end{displaymath}

\noindent They cannot be equal to one another for mixed events.
That means {\em there is some (non-dynamical) correlation between multiplicities
of kaons and pions in the mixed events} though the mixed events are constructed by
randomly choosing one particle from each randomly chosen real event. The correlation is
caused by the demand that the mixed events have the same total  multiplicity distribution as
the real events.

Despite such an intrinsic correlation between multiplicities of kaons and pions in the
mixed events, we can go further to analyze the $K/\pi$ ratio distribution in the those events.

Let
\begin{equation}\label{Eq:gnb}
\sum_N z^N M(N)=f(z)\ ,
\end{equation}

\noindent and using Eqs. (\ref{Eq:ffp}) and (\ref{Eq:ffg}),
the generating function for $P_{DF}(N,k)$ can be written as
\begin{equation}
\sum_k\sum_\pi x^ky^\pi P_{DF}(k+\pi,k)=f\left({Rx+y\over 1+R}\right) \ .
\end{equation}

It is clear that the ratio distribution in the mixed events depends on the shape
of total multiplicity distribution $f(z)$ and the mean ratio $R$.
In many experiments the multiplicity distribution can be parameterized as a
negative binomial (NB) or modified negative binomial (MNB) distributions, whose
generating functions are, respectively,
\begin{eqnarray*}
 G_{\rm NB}(z)=\left(1-r(z-1)\right)^{-\kappa}\\
G_{\rm MNB}(z)=\left({1-\Delta(z-1)\over 1-r(z-1)}\right)^\kappa
\end{eqnarray*}
\begin{figure}\centering
\includegraphics[width=0.5\textwidth]{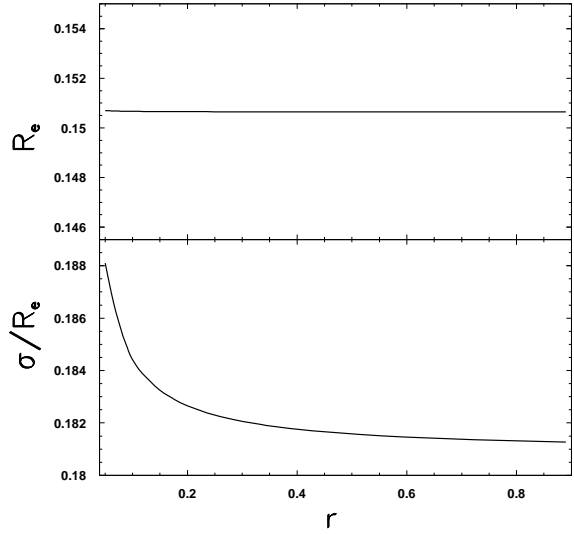}
\caption{Mean values $R_e$ and relative widths $\sigma/R_e$ of EBE $K/\pi$
ratio distributions as functions of parameter $r$ in the Negative Binomial
multiplicity distribution for mixed events.}
\label{Fig:4}
\end{figure}

The NB distribution can be regarded as a special case of the MNB distribution with
parameter $\Delta=0$. To see the dependence of the ratio distribution in the mixed events
on the shape of total multiplicity distribution or
on the parameters in the generating functions, we fix $R=0.15$ and the mean multiplicity
$\langle N\rangle=270.13$, the same as in \cite{gr97,fs99}. For the NB distribution,
$\langle N\rangle=r\kappa$, therefore after $\langle N\rangle$ is fixed only one
parameter is left which can be used
to specify the variance of total multiplicity distribution. In our calculation we choose
$r$ to change from 0.05 to 0.89. The calculated mean value and relative width
of the ratio distribution in the mixed events are shown in Fig. \ref{Fig:4}.
The mean value seems independent of the value of parameter $r$, while the width
decreases with $r$ slightly. Since the relative variance of the multiplicity distribution
has a linear dependence on the parameter $r$ in the NB distribution, the dependence
of the ratio $\sigma/R_e$ on the relative variance of the multiplicity distribution is similar
to that shown in \ref{Fig:4}.

For the MNB distribution, there are three parameters $\kappa, r$, and $\Delta$. In our
calculation, we fix $R=0.15$ and the mean multiplicity $\langle N\rangle=\kappa(r-\Delta)$
=270.13, the same as in the case for NB distribution, and we choose the relative variance of the
multiplicity distribution $w=(\langle N^2\rangle-\langle N\rangle^2)/\langle N\rangle=r+\Delta+1$
to be 1.0, 2.0, 2.2, and 2.5. Then the left parameter $\kappa$ is changed from 100 to 300.
The results for the mean values and the relative widths of the ratio
distribution are shown in Fig. \ref{Fig:5}. The mean $R_e$ is the same constant for all the
choice of parameters, while the relative width $\sigma/R_e$ is independent of $\kappa$ and
increases a little with the variance $w$ of the multiplicity distribution. The $w$ dependence
of the relative width $\sigma/R_e$ is a direct consequence of the multiplicity $N$ dependence
shown in Fig. \ref{Fig:2}. An important observation is that the calculated width $\sigma/R_e$
for the mixed events is much smaller than 23\% given in \cite{gr97,fs99,croland99}.
\begin{figure}\vskip -0.5cm\centering
\includegraphics[width=0.5\textwidth]{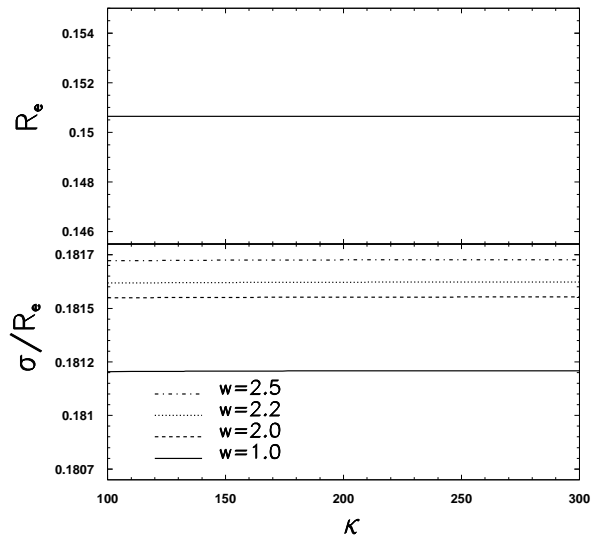}
\caption{Mean values $R_e$ and relative widths $\sigma/R_e$ of EBE $K/\pi$ ratio
distributions as functions of parameter $\kappa$ in the modified negative binomial multiplicity
distribution. Four variances $w$ for the multiplicity distribution are chosen for illustration.}
\label{Fig:5}
\end{figure}

\section{Mixing of events with {\em distributed} multiplicity and {\em distributed}
$K/\pi$ ratio}\label{Sec:dd}

The last and the most general case for our investigation of the mixing of real events is that
the real events have both the total multiplicity and $K/\pi$ ratio distributions (DD).
For given total multiplicity distribution $M(N)$
the corresponding joint probability for a mixed event to have $k$ kaons and $N-k$ pions is
\begin{equation}\label{Eq:ddp}
P_{\rm DD}(N,k)=M(N)P_{\rm FD}(N,k)\ .
\end{equation}

Because of the result in Sect. \ref{Sec:fd} that the ratio distributions in the mixed events
for cases FF and FD are almost the same, one has reason to expect that the results from
Eq.(\ref{Eq:ddp}) will be almost the same as those from Eq. (\ref{Eq:dfp}). So, this case
will not be discussed any more.

\section{Effect of imperfect identification on event mixing}
\label{Sec:intr}

In high energy heavy ion collisions, the identity of a particle is obtained from its charge and
$dE/dx$ information with the help of the Bethe-Bloch function. The rate of energy loss
$dE/dx$ depends on the charge, mass and momentum of the particle in a complex way, and
different particles may have very close $dE/dx$ values in some momentum range.
Therefore, normally one cannot make
definite conclusion about a particle's identity from such information. Instead one can only say,
for example, that a track is a pion or a kaon with a probability less than 1.
This means that a track which is thought to be a kaon may in fact
be a pion, and vise versa. If one tries to get $K/\pi$ ratio with the Maximum-Likelihood Method
from the $dE/dx$ information, as done in \cite{croland99}, such uncertainty will play a role
in the $K/\pi$ ratio distribution. Since the efficiency of identification depends on the
momenta of particles, it is extremely difficult to model the identification process.
In the following we only show that the identification efficiency
has important effect on the observed $K/\pi$ ratio distribution. To do that, we assume that
the identification efficiency is the same for all kaons (or pions).

For simplicity, we assume that only kaons and pions exist in the final states. Considering
the identification uncertainty of a track from its $dE/dx$ information, we assume
that the probability for a thought-to-be kaon to be indeed a kaon is $p_1$, and that for a
thought-to-be pion to be a real pion is $p_2$. We assume that the values of $p_1$ ($p_2$)
is the same for all thought-to-be kaons (pions). For an event with $K$
thought-to-be kaons and $N-K$ thought-to-be pions, the apparent $K/\pi$ ratio
is $R=K/(N-K)$. But the real numbers of kaons and pions may change from event to event
even if $N$ and $K$ are the same for all events, and the probability $P(N, k)$
of having $k$ real kaons and $N-k$ real pions can be calculated. In such a case, the $K/\pi$
ratio will have an intrinsic non-trivial distribution, although it is assumed
that every original
event has the same numbers of thought-to-be kaons and pions. It is not difficult to prove that
the generating function of the intrinsic distribution $P(N, k)$ is
\begin{equation}\label{Eq:intr}
 G^{N, K}_{\rm intrinsic}(x,y)=\left(p_1x+(1-p_1)y\right)^K\left((1-p_2)x+p_2y\right)^{N-K}\quad.
\end{equation}
\begin{figure}\vskip -0.5cm\centering
\includegraphics[width=0.5\textwidth]{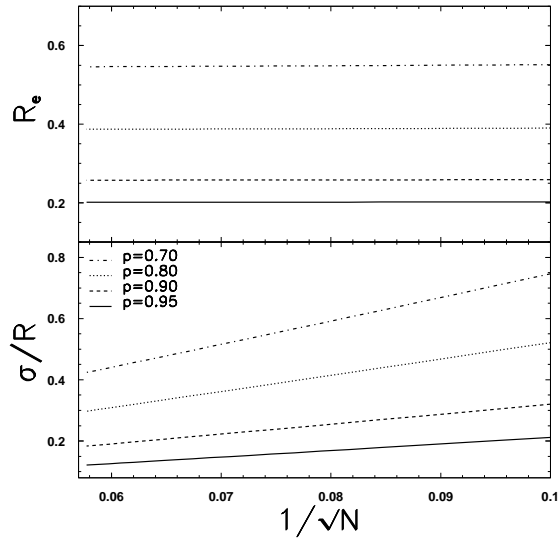}
\caption{Mean values $R_e$ and relative widths $\sigma/R$ of the intrinsic $K/\pi$ ratio
distribution as functions of the inverse root of multiplicity for
events without perfect particle identification. No event mixing is done here.}
\label{Fig:6}
\end{figure}

From this expression, we know that the real global $K/\pi$ ratio should be
\begin{equation}
R_g\equiv {\langle k\rangle\over \langle N-k\rangle}={p_1R+(1-p_2)\over
(1-p_1)R+p_2}\ ,
\end{equation}

\noindent which may be very different from the apparent ratio $R=K/(N-K)$. In the limiting case
$p_1=p_2=1/2$, i.e., when the kaons and pions can't be distinguished, the
global ratio $R_g$ will be 1, no matter how large $R$ is. Of course,
the global ratio is equal to the apparent one if $p_1=p_2=1$, as demanded physically.

We are particularly interested in the width $\sigma$ of the intrinsic $K/\pi$ ratio
distribution relative to the apparent mean ratio $R$, $\sigma/R$.
In the following we discuss the case with $p_1=p_2=p$. More general cases can be
studied in the same way. For definiteness we take in the following the apparent ratio
$R=0.15$, close to the $K/\pi$ ratios observed in current high energy heavy ion experiments.
The mean and relative width of the intrinsic ratio distribution are shown in
Fig. \ref{Fig:6} as functions of $1/\sqrt{N}$, inverse root of the event multiplicity.
This relative width calculated from Eqs. (\ref{Eq:intr}), (\ref{Eq:gl1}), and (\ref{Eq:gl2})
increases with the $1/\sqrt{N}$. Though linear relation between $\sigma/R$ and $1/\sqrt{N}$
can still be seen, the relation $\sigma\propto 1/\sqrt{N}$ is not longer a good approximation.
For poorer identification (smaller $p$) the relative width is larger for the same $N$.

Then we study the effect of event mixing on the $K/\pi$ ratio for the case with imperfect
identification. As in Sect. \ref{Sec:ff} we first fix the total multiplicity $N$
and the apparent $K/\pi$ ratio $R$. Using Eqs. (\ref{Eq:ffp}) and (\ref{Eq:intr}), one
can get the generating function for the joint distribution of real
kaon and pion multiplicities in the mixed events as
\begin{equation}\label{Eq:intrg}
 G_{\rm mixed}^{N, R}(x,y)=\left[{Rp+(1-p)\over 1+R}x+{p+R(1-p)\over 1+R}y\right]^N\ ,
\end{equation}

\noindent and the resulting mean values and relative width of the $K/\pi$ ratio distribution
are shown in Fig. \ref{Fig:7} as functions of inverse root of the total multiplicity $N$ for
several identification efficiency $p$. The mean values are again constants as in the case
without the event mixing, and the relative width decreases with $\sqrt{N}$ but is larger
than in Fig. \ref{Fig:6} for the same $\sqrt{N}$.
As in Fig. \ref{Fig:6} the relation $\sigma\propto 1/\sqrt{N}$ is neither a very good
approximation though the curves are almost linear. An important observation is that
$\sigma/R$ depends now strongly on the identification efficiency $p$ and can be about 23\%
for $p$ a little less than perfect value 1.0.
\begin{figure}\centering
\includegraphics[width=0.5\textwidth]{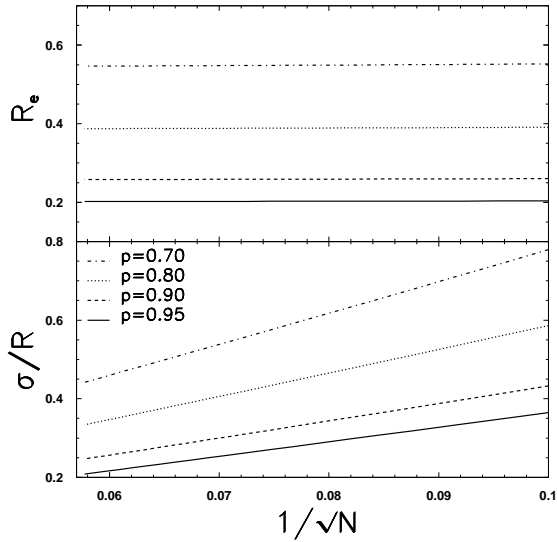}
\caption{Mean values $R_e$ and relative widths $\sigma/R$ as functions of inverse root of
the multiplicity for mixed events without perfect identification.}
\label{Fig:7}
\end{figure}

Now we can study the mixing of events with a multiplicity distribution and imperfect
particle identification. The generating function for the joint kaon and pion multiplicity
distribution in the mixed events can now be formulated using
Eqs. ({\ref{Eq:intrg}) and (\ref{Eq:gnb}). When the multiplicity distribution is assumed to
be a modified negative binomial distribution, the generating function for the joint kaon
and pion multiplicity distribution in the mixed events with imperfect particle
identification efficiency $p$ is
\begin{equation}
G_{\rm mixed}^{\rm MNB}(x,y)=\left({1-\Delta u\over 1-ru}\right)^\kappa\quad\quad \text{with}
\quad u={[Rp+(1-p)]x+[p+R(1-p)]y\over 1+R}\ .
\end{equation}

With this generating function we can perform theoretical calculation.
 We fix the mean multiplicity $\langle
N\rangle=\kappa(r-\Delta)=270.13$ as before. From Fig. \ref{Fig:5} we have seen that the ratio
distribution has extremely weak dependence on parameter $\kappa$. So that we fixed it to be 300.
Then we can change the relative variance, $w=(\langle N^2\rangle-\langle N\rangle^2)
/\langle N\rangle=1+r+\Delta$, of the total multiplicity distribution as a free parameter
from 1.9 to 2.9. Since $R_e$ in this case is neither measurable nor meaningful,
we focus only on the relative width $\sigma/R$ in the following.
The dependence of the relative width of the ratio distribution is shown in Fig. \ref{Fig:8}.
The relative width of the $K/\pi$ ratio distribution increases slowly with $w$
but has a very strong dependence on the identification efficiency $p$. For poorer
identification the relative width is larger. The value of the relative width
can be close to 23\% given in \cite{croland99} for quite high identification efficiency.
For other total multiplicity distributions, the mean and relative width of the EBE
$K/\pi$ ration distribution in the mixed events can also be calculated in a similar way.
\begin{figure}\centering
\includegraphics[width=0.5\textwidth]{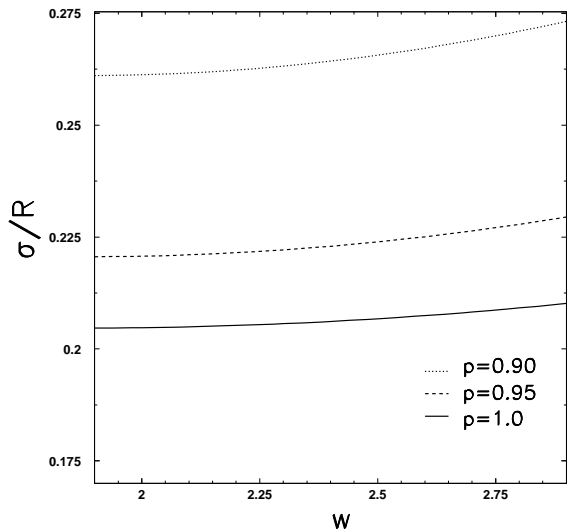}
\caption{Relative widths $\sigma/R$ as functions of $w$, which is
a measure of the variance of multiplicity distribution, for mixed events with a modified
negative binomial multiplicity distribution and without perfect particle identification.}
\label{Fig:8}
\end{figure}

\section{Discussion and conclusion}
\label{Sec:con}

From above sections one can know how to derive the EBE $K/\pi$ ratio distribution
for the mixed events once the total multiplicity distribution and global $K/\pi$
ratio in the real events are known. Then we can understand why the event mixing
method can be used to compare with results on $K/\pi$
ratio distribution from real events. Contrary to naive expectation,
there may exist correlation between multiplicities of kaons and pions in the mixed events.
Although the mean EBE $K/\pi$ ratio in the mixed events is stable for all cases with perfect
particle identification, the width shows a quite strong dependence on the total multiplicity
distribution and the global $K/\pi$ ratio. But the ratio distribution in the mixed
events has an extremely weak dependence on the shape of EBE $K/\pi$ ratio distribution
in the real events. This property enables us to estimate the dynamical contributions
to the ratio distribution in the real events. For events without perfect particle
identification, the intrinsic uncertainty of particle identity makes the relative width
of the $K/\pi$ ratio distribution in the mixed events larger and dependent on the
multiplicity distribution in a more complicated way. Currently observed relative width of the
EBE $K/\pi$ ratio distribution seems to be a combinational result from statistical fluctuations
(mainly due to the finite total multiplicity) and the imperfect particle identification.
More theoretical
and experimental studies are needed.

\acknowledgments
This work was supported in part by NNSF in China. Part of the work was done when one of
the authors (C.B.Y) was working in Max-Planck-Institut f\"ur Physik, Munich, Germany,
and he would like to thank the Alexander von Humboldt Foundation of Germany for the
financial support granted to him. He would also like to thank Dr. P. Seyboth for the
hospitality extended him during his stay in Munich. He also would like to thank
Drs. St. Mrowczynski and M. Gazdzicki for reading the draft and valuable comments.

\end{document}